# TURBULENT COOLING FLOWS IN MOLECULAR CLOUDS

P. C. Myers[1] AND A. Lazarian[2]

## ABSTRACT

We propose that inward, subsonic flows arise from the local dissipation of turbulent motions in molecular clouds. Such "turbulent cooling flows" may account for recent observations of spatially extended inward motions toward dense cores. These pressure-driven flows may arise from various types of turbulence and dissipation mechanisms. For the example of MHD waves and turbulence damped by ion-neutral friction, sustained cooling flow requires that the outer gas be sufficiently turbulent, that the inner gas have marginal field-neutral coupling, and that this coupling decrease sufficiently rapidly with increasing density. These conditions are most likely met at the transition between outer regions ionized primarily by UV photons and inner regions ionized primarily by cosmic rays. If so, turbulent cooling flows can help form dense cores, with speeds faster than expected for ambipolar diffusion. Such motions could reduce the time needed for dense core formation and could precede and enhance the motions of star-forming gravitational infall.

*Subject headings:* ISM: clouds — MHD — stars: formation — turbulence

## 1. INTRODUCTION

Recent maps of molecular cloud cores indicate inward motions that appear too extended to match models of gravitational infall and too fast to match models of gravitational contraction limited by ambipolar diffusion. In L1544 (Tafalla et al. 1998) and other "starless" cores (Lee, Tafalla, & Myers 1998) and in some "turbulent" cores with embedded stars (Mardones et al. 1998), a common feature is a zone of radius 0.05–0.1 pc, where most of the observed spectra have "infall asymmetry"—the skewing of an optically thick line to the blue of an optically thin line in a region with inward motions and centrally elevated excitation temperature (Leung & Brown 1977). This zone extends beyond the dense core as traced by the half-maximum contour of integrated intensity of the 1–0 line of $N_2H^+$, suggesting that the motions are associated with the formation of the core.

For L1544, a detailed analysis of the spectral profiles (Tafalla et al. 1998) indicates that the observations do not match the properties of "inside-out" gravitational collapse (Shu 1977). In that model, a star of luminosity several $L_\odot$ should have formed, whereas no star more luminous than 0.1 $L_\odot$ is known. The observed speeds, ~0.1 km s$^{-1}$, do not match the gravitational collapse of a uniform sphere from rest (Spitzer 1978), which predicts much less condensation than is observed. The inward motions may be inconsistent with ambipolar diffusion (Mestel & Spitzer 1956; Mouschovias 1976) since they are seen in both ionic and neutral lines and since their speed at 0.1 pc exceeds that predicted by the model "UV B" of Ciolek & Mouschovias (1995) by a factor of ~5.

These comparisons do not rule out all versions of gravitational infall or all versions of gravitational contraction limited by ambipolar diffusion (e.g., Safier, McKee, & Stahler 1997; Basu 1998; Li 1998). Nonetheless, they suggest that nongravitational motions should also be considered as ways to form dense cores. In this Letter, we propose that some extended inward motions reflect pressure-driven cooling flows associated with local dissipation of turbulence. In a related paper, we consider additional mechanisms of inward motions (Lazarian & Myers 1998).

## 2. TURBULENT COOLING FLOWS

"Turbulent cooling flows" may follow localized dissipation of turbulence in molecular clouds, as "cooling flows" follow radiative cooling in clusters of galaxies (e.g., Cowie & Binney 1977). Molecular cloud turbulence likely has a substantial MHD component (Arons & Max 1975), since the Alfvén speed is comparable to the nonthermal component of the velocity dispersion (Myers & Goodman 1988; Crutcher 1998). MHD waves and turbulence can dissipate due to wave damping by ion-neutral friction (Kulsrud & Pearce 1969), nonlinear wave steepening (Cohen & Kulsrud 1974; Zweibel & Josafatsson 1983), and interactions of nonlinear Alfvén waves (Kraichnan 1965). MHD waves excited during core formation are expected to dissipate due to ion-neutral friction within a few core free-fall times unless new waves are internally excited, e.g., by winds from young stars (Nakano 1998). Compressible MHD turbulence decays nearly as fast as nonmagnetic hydrodynamic turbulence, in little more than a free-fall time, according to simulations (Ostriker, Gammie, & Stone 1998; MacLow et al. 1998). A turbulent cooling instability may arise as density increases, reducing the field-neutral coupling and velocity dispersion, causing an inward pressure gradient and flow, increasing the density further (Myers & Khersonsky 1995). Thermal cooling in molecular clouds (Meerson, Megged, & Tajima 1996) is closely related, as are cooling to form clouds (Elmegreen 1989) and wave-driven flows in diffuse clouds (Elmegreen 1997).

We present a simple model to specify the conditions that allow a turbulent cooling flow. We assume a uniform outer region of density $\rho_{\rm out}$ and a uniform spherical inner region of radius $r$ and density $\rho_{\rm in}$. The regions have common isothermal sound speed $\sigma_T$ and common Alfvén speed $v_A$, in accord with the tendency for magnetic field strength to increase as $\rho^{1/2}$ (Crutcher 1998). The outer and inner pressures on the boundary surface normal to the field are respectively $P_{\rm out} = \rho_{\rm out}(\sigma_T^2 + \sigma_{\rm NT,out}^2)$ and $P_{\rm in} = \rho_{\rm in}(\sigma_T^2 + \sigma_{\rm NT,in}^2)$, where $\sigma_{\rm NT}$ denotes the nonthermal component of the velocity dispersion. Each region can propagate MHD waves, and the outer region has sufficient stars and stellar winds to maintain waves with equipartition energy density, so that $\sigma^2_{NT,out} = (v_A^2)/3$. The inner region has no internal sources of wave excitation, so its turbulence arises solely from externally incident waves. The inner turbulence is sufficiently "wavelike" that it has negligible shocks and nonlinear interactions among its waves. A similar model of MHD wave pressure is presented by Martin, Heyvaerts, & Priest (1997).

[1] Harvard-Smithsonian Center for Astrophysics, 60 Garden Street, Cambridge, MA 02138; pmyers@cfa.harvard.edu.
[2] Princeton University Observatory, Peyton Hall, Princeton, NJ 08544; lazarian@astro.princeton.edu.

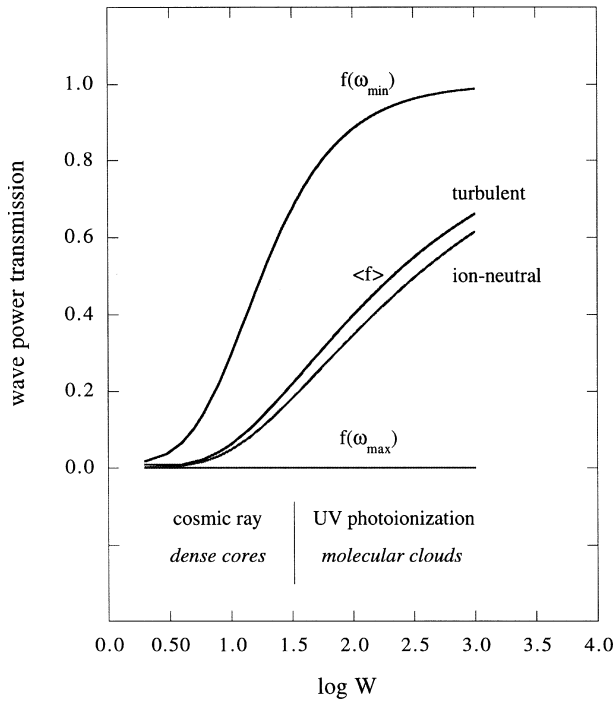

Fig. 1.—MHD wave power transmission $f$ into a uniform spherical region as a function of $W$, the ratio of radius to MHD cutoff wavelength. The upper and lower curves represent monochromatic transmission at the lowest and highest frequencies of wave propagation in the region. The middle curves are averages over a wave energy spectrum $\sim \omega^{-3/2}$ as for the inertial range of uncorrelated MHD turbulence, assuming that $\omega_{max}$ is set either by the end of the turbulent inertial range or by the critical frequency for ion-neutral damping. These averages $\langle f \rangle$ indicate significant damping of MHD waves in dense cores (ionized by cosmic rays) and in their surrounding molecular clouds (ionized primarily by UV photons).

The net inward pressure is $\Delta P \equiv P_{out} - P_{in}$, or from the above equations

$$\Delta P = [\rho_{out}\sigma_T^2][1 - \eta + (x^2/3)(1 - \eta g)], \quad (1)$$

where $\eta \equiv \rho_{in}/\rho_{out}$, $x \equiv v_A/\sigma_T$, and $g \equiv \sigma^2_{NT,in}/\sigma^2_{NT,out}$. If $\Delta P > 0$, the resulting inward flow has speed $v \approx [\Delta P/\rho_{out}]^{1/2}$, and consequently the flow has sonic Mach number

$$v/\sigma_T \approx [1 - \eta + (x^2/3)(1 - \eta g)]^{1/2}. \quad (2)$$

From equation (2), the density contrast $\eta$ reaches its maximum, $1 + x^2/3$, when $v = 0$ and $\eta g \ll 1$.

To obtain the nonthermal velocity variance ratio $g$, we assume an isotropic flux of externally incident waves, which have negligible reflection at the boundary. We take the spatial mean of the monochromatic wave power transmission from the surface to the inside of the sphere as $f \equiv e^{-u}$, where $u \equiv 2\kappa r$ and $\kappa$ is the imaginary part of the uniform propagation constant. We checked that $f$ is a good enough approximation to $f' \equiv 3u^{-2}e^{-u}[(1 + 3u^{-1})\cosh(u) - (1 + u^{-1} + 3u^{-2})\sinh(u)]$, the proper average over all radii and angles, and found that $f$ lies within 15% of $f'$ for $0 \leq u \leq 3$.

So far, this model applies to any kind of turbulent motions that exert pressure and to any dissipation mechanism. To make specific predictions, we consider the example of dissipation due to wave damping by ion-neutral friction, since this mode of turbulent dissipation appears likely to dominate over shocks and nonlinear mode coupling in the relatively quiescent regions in which extended inward motions are currently observed. We evaluate $\kappa$ from the dispersion relation for waves of angular frequency $\omega$ in a lightly ionized magnetized gas (Kulsrud & Pearce 1969):

$$\kappa = \frac{\omega}{v_A[1 + (\omega\tau)^2]^{1/4}} \sin\left[\frac{1}{2}\arctan(\omega\tau)\right], \quad (3)$$

where $\tau$ is the neutral-ion collision time. The angular frequency ranges from $\omega_{min} \approx 2\pi v_A/r$, which corresponds to the longest wavelength wave that can propagate in the inner region ($\lambda = r$), to the cutoff frequency $\omega_{max}$. We use two values that probably bracket $\omega_{max} - \omega_{max}$(turbulent), corresponding to the high-frequency end of the inertial range for MHD turbulence, beyond which the spectrum decays exponentially (e.g., Biskamp 1993), and $\omega_{max}$(ion-neutral), where the wavelength equals the ion-neutral cutoff wavelength $\lambda_0$ (Kulsrud & Pearce 1969). In terms of the field-neutral coupling parameter $W = r/\lambda_0$ (Myers & Khersonsky 1995), $\omega_{max}$(turbulent) $\approx \omega_{min}W^{2/3}$ and $\omega_{max}$(ion-neutral) $= \omega_{min}W$. We assume a wave energy spectrum $E(\omega) d\omega \sim \omega^{-3/2} d\omega$ as for the inertial range of uncorrelated MHD turbulence (Kraichnan 1965), so

$$g = \int_{\omega_{min}}^{\omega_{max}} d\omega \, \omega^{-3/2} \exp(-2\kappa r) \bigg/ \int_{\omega_{min}}^{\omega_{max}} d\omega \, \omega^{-3/2}. \quad (4)$$

From equation (3) and the adopted definitions of frequency limits, equation (4) can be expressed in terms of the single variable $W$. We evaluated equation (4) by numerical integration for $W \leq 10$. For larger $W$ we used the approximation $\omega\tau \ll 1$, which conveniently reduces equation (4) with sufficient accuracy to a combination of elementary functions and the incomplete gamma function.

Figure 1 shows that $g$ decreases with decreasing field-neutral coupling parameter $W$, depends sensitively on the frequency range, and falls below unity even for "marginal" coupling, i.e., $W < 100$. For example, when $W = 100$, $g = 0.9$ (negligible damping) if only the lowest frequency waves, with $\omega = \omega_{min}$, are considered; $g = 0.35$–$0.40$ (substantial damping) for the spectral averages in equation (4), and $g \approx 0$ (complete damping) if only the waves at either value of $\omega_{max}$ are considered. For the spectral averages, $g$ decreases from $\sim 1$ when $\log W > 3$, as for regions of density $n < 10^3$ cm$^{-3}$ (Myers & Khersonsky 1995), to $\sim 0.1$ when $W = 10$–$30$, as for dense cores with $n = 10^4$–$10^5$ cm$^{-3}$ (Williams et al. 1998; Bergin et al. 1998), to $\sim 0$ when $W = 1$ as for completely cutoff condensations. These results change only slightly if the wave energy spectrum varies as $\omega^{-5/3}$, as expected for hydrodynamic turbulence (Kolmogorov 1941) or "strong Alfvénic turbulence" (Goldreich & Sridhar 1995). These results support the idea that starless cores with low ionization should soon dissipate the turbulence associated with their formation (Nakano 1998).

Figure 2 shows requirements for "instantaneous" and "sustained" turbulent cooling flow. For our example of ion-neutral damping, the condition for flow at any moment, $\Delta P > 0$ in equation (1), implies that the coupling parameter $W$ be less than an upper limit $W_{max}$ given by equations (1) and (4) for specified values of $x$ and $\eta$. Figure 2 shows log $W_{max}$ as a function of $x$ for density contrast values $\eta = 1.5$, 2, and 3. In order to have instantaneous cooling flow into a region with substantial density contrast, e.g., $\eta = 2$, $W < W_{max}$ implies that the outer region must be turbulent ($x > \sim 2$) and the inner region must have "marginal" field-neutral coupling ($W < 100$). Further, if the flow is to be self-sustaining, $W$ must decrease sufficiently rapidly as the inner density increases, e.g., for $x = 3$, from $W < 300$ when $\eta = 1.5$ to $W < 10$ when $\eta = 3$.

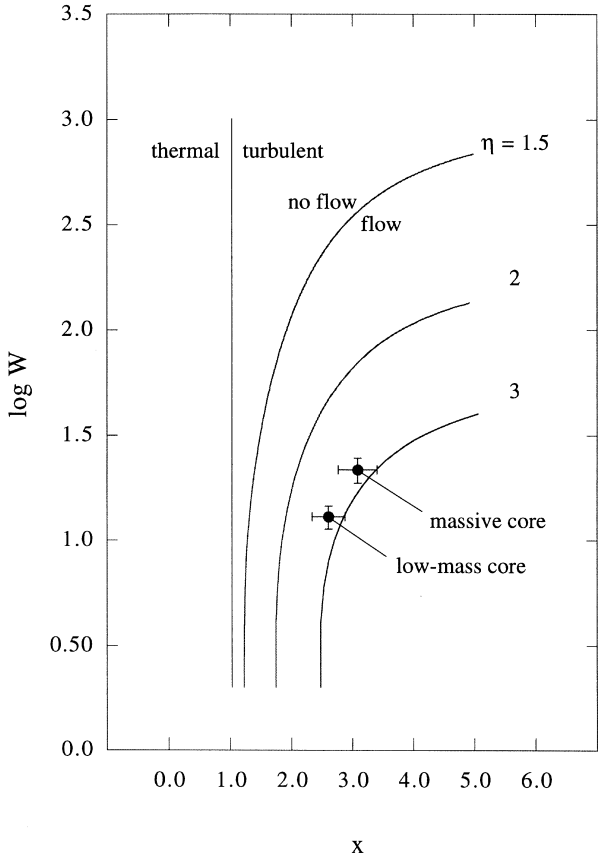

Fig. 2.—Critical conditions for turbulent cooling flow. For each value of the density contrast $\eta$ between inner and outer regions and of the Alfvén Mach number $x$, a turbulent cooling flow can exist only when the field-neutral coupling parameter $W$ in the inner region is less than the critical curve shown. Thus, instantaneous flow requires significant turbulence in the outer region ($x > 1$) and marginal coupling in the inner region ($W < 100$), and sustained flow also requires sufficient decrease in coupling with increasing density. The data points indicate mean values of $W$ and $x$ for samples of low-mass and more massive cores, and each error bar represents the standard error of the mean. These data are consistent with cooling flows onto dense cores.

A likely site for such a sustained turbulent cooling flow is an outer region dominated by UV ionization and an inner region dominated by cosmic-ray ionization, with the boundary at visual extinction $A_{v,0} \approx 4$ mag from a cloud surface illuminated by the standard interstellar radiation field (McKee 1989). This is so because the inward pressure in equation (1) is greatest at radii $\sim 0.1$ pc for low-mass cores, according to the ionization model of McKee (1989) and the density model of Caselli & Myers (1995), as in Myers (1997). This radius is similar to that of observed infall asymmetry. At larger radii, the coupling $W$ is probably too strong for significant flow. At smaller radii, $W$ is weak enough for flow to arise, but $W$ may not decrease fast enough with increasing density to sustain the flow. These conjectures need more detailed models for confirmation.

Figure 2 also indicates that observational estimates of $W$ and $x$ are consistent with requirements for turbulent cooling flows onto the typical dense core. Each of the two data points represents a mean value of $W$, according to determination of electron abundance from observations of $H^{13}CO^+$, $DCO^+$, and $C^{18}O$ and from chemical models of neutral and ion abundance, for samples of 20 low-mass cores (Williams et al. 1998) and seven "massive" cores (Bergin et al. 1998). The value of $x$ is based on assuming that the cores have kinetic temperatures 10 and 15 K, respectively, and that the velocity dispersion in the outer region is represented by the mean line width of $C^{18}O$, which traces lower density gas than do the lines of $H^{13}CO^+$ and $DCO^+$.

The mean density contrast $\eta = 2$–3, which is consistent with the models and data in Figure 2, is also comparable to that expected for gravitationally bound condensations. The mean density over a spherically symmetric region of radius $r$ is greater than the density at $r$ by a factor of 2.44 for a critically stable Bonnor-Ebert sphere (Bonnor 1956; Ebert 1955) and by a factor of 3 for a singular isothermal sphere (Shu 1977). Thus, cooling flow may help condense gas to the point at which it is self-gravitating and may then accompany the ensuing gravitational infall. If so, cooling flows can modify and enhance the process of gravitational infall.

The characteristic speeds of inflow and boundary expansion are similar to the sound speed. We assume $x = 3$ as in a region whose line widths are dominated by turbulent motions and that $\eta$ has reached half of its maximum value, i.e., $\eta = 2$. Then, for $W = 20$ as in many dense cores, equations (2) and (4) give $v = 1.0\sigma_T$, assuming that $\omega_{max} = \omega$(turbulent). Similarly, the outward speed of the boundary between the inner and outer regions $dr/dt$ is estimated from the continuity equation if $\eta$ is constant as assumed above: $dr/dt = v/\eta$, or for $x = 3$, $\eta = 2$, and $W = 20$, $dr/dt = 0.5\sigma_T$. For 10 K gas, the condensation radius would grow from 0.05 to 0.10 pc in $\sim 0.5$ Myr.

## 3. DISCUSSION

The pressure-driven flow described here may be an effective way to condense molecular cloud gas. Such flows may occur wherever turbulent motions dissipate locally, with flow speed less than the effective sound speed of the outer gas. For wavelike MHD turbulence with ion-neutral damping and a $\omega^{-3/2}$ spectrum, the flow speed is $\sim 0.1$ km s$^{-1}$, substantially faster than generally expected for ambipolar diffusion. If such a flow operates for 0.1–1 Myr, it may bring more mass into a dense core than does ambipolar diffusion, thus reducing the time of core condensation. Shorter core formation times would help account for the relatively low proportion of starless dense cores observed in Taurus and Orion, which suggest that starless cores are detectable for less than the few Myr expected from ambipolar diffusion (Jijina, Myers, & Adams 1998).

Turbulent cooling flows should also bear on the line width–size relations, in which nonthermal line widths increase with map size and decrease with mean map density (Larson 1981; Myers & Goodman 1988; Falgarone, Puget, & Pérault 1992; Goodman et al. 1998). Since turbulent cooling flows increase the density of gas with low-velocity dispersion, they should tend to decrease the line width more on small size scales than on large size scales. If such flows are sufficiently prevalent, they should contribute to the trends evident in the line width–size relations.

On the other hand, much more work is needed to fully describe the flows outlined here and to compare their properties with observations. Our treatment ignores the role of the static magnetic field in retarding inward motion and thus requires that the flow be primarily along field lines. It neglects the excitation of turbulence in the inner region by the flow and exaggerates the flow pressure gradient by considering a uniform inner region and equipartition turbulence in the outer region. We consider only one type of "turbulence," a linear superposition of MHD waves, and only one mechanism of dissipation, ion-neutral friction, while wave steepening and nonlinear effects should also be important. We ignore the role of self-gravity, while in reality turbulent cooling might precede and enhance gravitational infall. Finally, it is desirable to provide a space-time description of flow density and velocity to allow detailed comparison with observations. These topics are suitable for future papers, and some are discussed by Lazarian & Myers (1998).

We thank Fred Adams, Bruce Draine, Bruce Elmegreen, Charles Gammie, Alyssa Goodman, and Mario Tafalla for helpful comments and Frank Shu for detailed comments and suggestions.